\def\re#1{(\ref{#1})}
\def\beq{\begin{equation}}
\def\eeq{\end{equation}}
\def\beeq{\begin{eqnarray}}
\def\beeqn{\begin{eqnarray*}}
\def\eeeq{\end{eqnarray}}
\def\eeeqn{\end{eqnarray*}}
\def\nome#1{{\label{#1}}}
\def\ln{large-$N\;$}

\def\th{\theta}

\renewcommand{\AA}{{\cal A}}

\newcommand{\OO}{{\cal O}}

\newcommand{\ZZ}{{\cal Z}}
\newcommand{\WW}{{\cal W}}
\newcommand{\Wn}{{\cal W}_n}

\newcommand{\lp}{\left(}
\newcommand{\rp}{\right)}
\renewcommand{\lq}{\left[}
\renewcommand{\rq}{\right]}

\newcommand{\no}{\nonumber}
\newcommand{\ph}{\phantom}

\def\frac#1#2{ {{#1} \over {#2} }}

\def\ie{\hbox{\it i.e.}{ }}

\def\bom#1{\mbox{\boldmath$#1$}}
\def\en{\bom{n}}

\documentstyle[preprint,aps,amssymb]{revtex}
\begin{document}
\draft
\newfont{\form}{cmss10}

\title{Two-dimensional QCD and instanton contribution~\footnote{
Talk given by F. Vian at the {\it
XVII Autumn School}, Lisbon (Portugal), September 29 - October 4, 1999.}}

\author{A. Bassetto$^1$, L. Griguolo$^2$ and
F. Vian$^1$} 
\address{$^1$Dipartimento di Fisica "G. Galilei",
INFN, Sezione di Padova,\\
Via Marzolo 8, 35131 Padua, Italy\\
$^2$Dipartimento di Fisica ``M. Melloni'',
INFN, Gruppo Collegato di Parma, \\ Viale delle Scienze, 43100 Parma, Italy}

\maketitle

\begin{abstract}

In these notes Yang-Mills theories in $1+1$
dimensions are reviewed. Instantons on a sphere prove to be ---in the
decompactification limit--- the key issue to clarify an old
controversy between equal-time and  light-front quantization.

\end{abstract}

\vskip 2.0truecm
\noindent
DFPD 99/TH 49

\noindent
UPRF-99-18

\vfill\eject

\narrowtext

\section{Introduction}
In the past few years a large amount of efforts has been devoted to
two-dimensional gauge theories ($YM_2$), mainly motivated by the
possibility of exact solutions which are believed to share features
with the real four-dimensional world.

Although $YM_2$ seems  trivial when quantized
in the light-cone gauge, 
still topological degrees of freedom occur if the theory is put
on a (partially or totally) compact manifold, whereas the simpler 
behaviour enforced by the light-cone gauge condition on the plane
entails a severe worsening in its infrared structure.
One can say that, in light-cone gauge, dynamics gets hidden
in the very singular nature of correlators at large distances
(IR singularities). 

The first quantity that comes to mind is the 
two-point correlator. If the theory is quantized in the gauge
$A_{-}=0$ at {\em equal-times}, the free propagator has the following
causal expression (WML prescription) in two dimensions
\begin{equation}
\label{WMLprop}
D^{WML}_{++}(x)={1\over {2\pi}}\,\frac{x^{-}}{-x^{+}+i\epsilon x^{-}}\,,
\qquad\qquad x^{\pm}=\frac{x^0\pm x^1}{\sqrt2},
\end{equation}
first proposed by T.T. Wu~\cite{wu}. In turn this propagator is nothing
but the restriction in two dimensions of the expression proposed 
by S. Mandelstam and G. Leibbrandt~\cite{mandel} in four dimensions and 
derived by means of a canonical quantization in ref.~\cite{bosco}.

In dimensions higher than two, where ``physical'' degrees of freedom
are switched on (transverse ``gluons''), 
causality is mandatory in order to
get correct analyticity properties, which in turn are the basis of
any consistent renormalization program, and to obtain
agreement 
with Feynman gauge results~\cite{pertu}.

The situation is somewhat different in exactly two dimensions. Here 
the theory can be quantized on the {\em light-front} (at equal $x^{+}$);
with such a choice, 
no dynamical degrees of freedom occur as the non-vanishing 
component of the vector field does not propagate
\begin{equation}
\label{CPVprop}
D^{P}_{++}(x)=-\frac{i}{2}|x^{-}|\,\delta(x^{+}),
\end{equation}
but rather gives rise to an instantaneous (in $x^{+}$) Coulomb-like
potential. 

A formulation based essentially on the potential in Eq.~\re{CPVprop}
was originally proposed by G. 't Hooft in 1974~\cite{hooft}, to derive
beautiful solutions for the $q\bar q$-bound state problem under the form of
rising Regge trajectories. On the other hand, when Wu's prescription
is adopted, the bound state equation at \ln turns out to be very
difficult, and Regge trajectories disappear. 
A suitable tool to clarify the origin of this discrepancy is the Wilson
loop, owing to its gauge invariance and to its reasonable infrared
properties.

\section{Wilson loop on the plane}

When inserted in perturbative Wilson loop calculations,
expressions Eqs.~\re{WMLprop} and ~\re{CPVprop} lead to completely
different results.
In two
dimensions pure-area exponentiation is expected. This is due to the
invariance of the theory under area-preserving diffeomorphisms, which
suggests that the Wilson loop is a function of the dimensionless
quantity $g^2 \AA$, with $\AA$ the encircled area, and to unitary
evolution.

If a rectangle with light-like sides is chosen as a contour, with 't
Hooft potential only planar diagrams can be built, for any value
of $N$. Therefore the perturbative series is easily resummed leading
to the expected result~\footnote{in the Euclidean formulation} 
\beq \nome{expo}
W[\AA]=\exp \lp -\frac12  C_F g^2 \AA \rp \,,
\eeq
where $C_F$ is  the quadratic Casimir operator for the fundamental
representation, \ie $\frac{N^2-1}{2N}$, $\frac{N}2$ for $SU(N)$,
$U(N)$, respectively.
Confinement holds and, in the 't Hooft limit $N \to \infty$, $g^2 N$
fixed, Eq.~\re{expo} exhibits a finite string tension.

When the propagator is endowed with the causal prescription
Eq.~\re{WMLprop}, instead, disagreement with the result
Eq.~\re{expo} already shows up at order $\OO (g^4)$, and is due to graphs
with crossed vector propagators which produce a contribution
proportional to $C_FC_A$, $C_A$ being the Casimir of the adjoint
representation. Although non-trivial, resummation of the perturbative
series at all orders in the coupling constant $g$ is still
possible. This was performed in~\cite{tedeschi}: once a circular
contour is chosen, geometry factorizes out and the task of determining
the Wilson loop reduces to the purely combinatorial problem of finding
the group factors corresponding to the Wick contractions. Fortunately
these factors are generated by a matrix integral in the space of
hermitian (traceless) $N\times N$ matrices for $U(N)$ ($SU(N)$). 
The result for
$U(N)$  reads
\begin{equation}
\label{laguerre}
\WW=\frac{1}{N}\exp\left[-\frac{g^2 \AA}4\right]\,
L_{N-1}^{(1)}(g^2 \AA/2)\,,
\end{equation}
the function $L_{N-1}^{(1)}$ being a generalized Laguerre
polynomial. In addition to the extra polynomial appearing in
Eq.~\re{laguerre}, the string tension turns out to be different from
the one in Eq.~\re{expo}. More dramatically, Eq.~\re{laguerre} in the 't
Hooft limit becomes 
\begin{equation}
\label{bessel}
\lim_{N\to \infty} \WW=
\sqrt{ \frac{2}{ {\hat{g}}^2 \AA} }
J_1 \Bigl(\sqrt{2{\hat{g}^2} \AA}\Bigr)\,,
\end{equation}
where ${\hat{g}}^2=g^2\,N$, and  confinement is lost, thereby explaining 
the failure of Wu's spectrum.

We are now standing before two different scenarios and there is
apparently no motivation to prefer one to the other, since both 't
Hooft's and Wu's results are analytic in $g^2$. The need for
reconciliation becomes urgent.

\section{The instanton expansion}

In order to gain a deeper insight, it is worthwhile to study the
problem on a compact two-dimensional manifold~\cite{gross}. Let it be
the sphere for simplicity. Therein  the procedure is purely
geometrical and group-theoretical, so that no gauge-fixing has to be
adopted and infra-red problems are absent.

The starting point are the well-known expressions~\cite{migdal} of the
exact partition function and of a Wilson loop for  a pure $U(N)$
Yang-Mills theory on a sphere $S^2$ with area $A$
\begin{eqnarray}
\label{partition}
{\cal Z}(A)&=&\sum_{R} (d_{R})^2 \exp\left[-{{g^2 A}\over 4}C_2(R)\right]\,,\\
\label{wilson}
\WW (A-{\cal A},{\cal A})&=&\frac1{\ZZ N}\sum_{R,S} d_{R} \,
d_{S} \exp\left[-{{g^2 (A-{\cal A})}\over 4}C_2(R)-{{g^2 {\cal
A}}\over 4}C_2(S)\right] \no \\ 
& \times  & \int dU {\rm Tr}\lq U \rq \chi_{R}(U) \chi_{S}^{\dagger}(U),
\end{eqnarray}
$d_{R\,(S)}$ being the dimension of the irreducible
representation $R(S)$ of $U(N)$; $C_2(R)$ ($C_2(S)$) is the quadratic
Casimir, $A-{\cal A},{\cal A}$ are the areas singled out by the loop,
the integral in Eq.~\re{wilson} is over the $U(N)$ group manifold and
$\chi_{R(S)}$ is the character of the group element $U$ in the
$R\,(S)$ representation. 
In order to evaluate $\cal W (A-{\cal A},{\cal A})$ in the
decompactification 
limit, for $N>1$ Eqs.~\re{partition}, \re{wilson} can be explicitly written in the form
\beeq
\label{partip}
&& {\cal Z}(A)=
\frac{1}{N!}\exp \left [ -\frac{g^2 A}{48}N(N^2-1)\right ]
\sum_{m_i = -\infty}^{+\infty}\Delta^2(m_1,...,m_N) \no \\
&& \ph{{\cal Z}(A) }
\times \exp\left [ 
-\frac{g^2A}{4}\sum_{i=1}^N \lp m_i-\frac{N-1}{2} \rp^2\right]\,,\\
\label{wilsonp}
&&\WW (A-\AA, \AA)=\frac1{\ZZ N N!}\exp \left[-\frac{g^2A}{48}N(N^2-1)
\right ] \no \\
&&\times\sum_{k= 1}^{N}\sum_{m_i=-\infty}^{+\infty}
\Delta(m_1+\delta_{1,k},...,m_N+\delta_{N,k}) 
\Delta(m_1,...,m_N)\\
&&\times
\exp\left [-\frac{g^2 (A-\AA)}{4}\sum_{i=1}^N \lp m_i-\frac{N-1}{2} \rp^2
 -\frac{g^2 {\cal A}}{4}\sum_{i=1}^N \lp
m_i-\frac{N-1}{2}+\delta_{i,k}
\rp^2\right]\,.
\nonumber
\end{eqnarray}
In the previous formulae the generic irreducible representation has
been described by means
of the set of integers $m_{i}=(m_1,...,m_{N})$, related to the
Young tableaux, in terms of which one  gets
\begin{eqnarray}
\label{casimiri}
C_2(R)&=&\frac{N}{12}(N^2-1)+\sum_{i=1}^{N}\lp m_{i}-\frac{N-1}{2}\rp^2,
\nonumber \\
d_{R}&=&\Delta(m_1,...,m_{N}).
\end{eqnarray}
$\Delta$ is the Vandermonde determinant and
 the integration in Eq.~(\ref{wilson})
has been performed explicitly, using the well-known formula for the 
characters in terms of the set $m_{i}$.

Now, as first noted by Witten~\cite{witte}, it is possible to
represent ${\cal Z}(A)$ ---and consequently $\WW (A-{\cal A},{\cal
A})$--- as a sum over instable instantons, each instanton contribution
being associated to a finite, non-trivial, perturbative expansion. One
can observe that in the sum over $\{m_i\}$ the dependence on the area
is through the dimensionless quantity $g^2 A$, whereas an instanton
action typically depends on $\frac1{g^2 A}$. Therefore a duality
transformation is required to  turn the fraction upside down. The
mathematical tool to carry out such a task is provided by a Poisson
resummation~\cite{case} over $\{m_i\}$
\begin{eqnarray}
\label{poisson}
&&\sum_{m_{i}=-\infty}^{+\infty}F(m_1,...,m_{N})=
\sum_{n_{i}=-\infty}^{+\infty}\tilde{F}(n_1,...,n_{N})\,,\\
&&\tilde{F}(n_1,...,n_{N})=\int_{-\infty}^{+\infty}dz_1...dz_{N}
\exp \left[2\pi i(z_1 n_1+...+z_{N}n_{N})\right]\,F(z_1,...,z_{N})\,. \no
\end{eqnarray}
When performed in Eqs.~\re{partip}, \re{wilsonp}, it gives
\begin{eqnarray}
\label{instant}
{\cal Z}(A)&=&C(g^2 A,N)\sum_{n_{i}=-\infty}^{+\infty}
\exp\left[-S_{inst}(n_{i})\right]Z(n_1,...,n_{N}),\nonumber\\
{\cal W}(A-\AA, \AA)&=&\frac{1}{{\cal Z}N}C(g^2 A,N)\exp \left[
-g^2\frac{A (A-\AA)}{4A}\right]\sum_{n_{i}=-\infty}^{+\infty}
\exp\left[-S_{inst}(n_{i})\right]\nonumber\\
&\times&\sum_{k=1}^{N}\exp\left[-2 \pi i n_{k}\frac{A-\AA}{A}\right]
W_{k}(n_1,...,n_{N}),
\end{eqnarray}
where
\beq
\label{semicl}
S_{inst}(n_{i})=\frac{4\pi^2}{g^2 A}\sum_{i=1}^{N}n_{i}^2\,.  \eeq
Finally, $S_{inst}(n_{i})$ is interpreted as an instanton
action~\footnote{Indeed, on $S^2$ there are non-trivial solutions of
the Yang-Mills equation, labelled by the set of integers
$\{n_i\}=(n_1, \ldots, n_N)$ (see~\cite{capo}).}, while $Z(n_1,...,n_{N})$ and
$W_{k}(n_1,...,n_{N})$ are the quantum corrections around the
classical solutions~\cite{capo}.

From the above representations it is rather clear why the
decompactification limit $A\to \infty$ should not be performed too
early. As a matter of fact, on the plane fluctuations around the
instanton solutions are undistinguishable from Gaussian fluctuations
around the trivial field configuration, since $S_{inst}(n_{i})$ goes
to zero for any finite set $n_{i}$ when $A\to \infty$. For finite $A$
and finite $n_{i}$ instead, in the limit $g\to 0$, only the
zero-instanton sector can survive in the Wilson loop expression.  

\section{Relation with perturbation theory}

In principle the zero-instanton contribution should be obtainable 
by means of perturbative calculations.  
If in Eqs.~\re{instant}, \re{semicl} only  the zero-instanton sector, \ie 
$n_{i}=0$, is retained, after some technicalities the following result
is found~\cite{capo}
\begin{equation}
\label{result}
{\cal W}^{(0)}=\frac{1}{N}\exp\left[-g^2\frac{\AA (A-\AA)}{4A}\right]\,
L_{N-1}^1 \left(g^2\frac{\AA (A-\AA)}{2A}\right)\,.
\end{equation}
At this point a remark is in order: in the decompactification limit $A\to
\infty$, $\AA$ fixed, the quantity in the equation above {\it exactly}
coincides, for any value of $N$, with Eq.~\re{laguerre}, which
was derived from a matrix model.
Hence, the zero-instanton contribution corresponds in a sense to
``integrating over the group algebra''. 

\section{The Wilson loop with winding number \en}

Disagreement of Eq.~\re{result} with the pure-area exponentiation is
at this stage no longer surprising since ${\cal W}^{(0)}$ does not contain
any genuine non-perturbative contribution, {\it viz} instantons.  For
any value of $N$ the pure-area exponentiation follows, after
decompactification, from resummation of all instanton sectors,
changing completely the zero-sector behaviour and, in particular, the
value of the string tension.

What might instead be unexpected in this context is the fact that, using the 
instantaneous 't Hooft potential and just
resumming at all orders the
related perturbative series, one still ends up with the
correct pure-area exponentiation. 

This is true also in more general instances, such as Wilson loops
winding $n$ times on a closed contour~\cite{bgv}. 
For a pure $U(N)$ Yang-Mills theory on a sphere $S^2$ with area $A$ it holds
\beeq
\label{wilsonn}
&& {\cal W}_n(A-{\cal A},{\cal A})={1\over {\cal Z}N}\sum_{R,S} d_{R}
\, d_{S} \exp\left[-{{g^2 (A-{\cal A})}\over 4}C_2(R)-{{g^2 {\cal
A}}\over 4}C_2(S)\right] \no \\ && \ph{{\cal W}_n(A-{\cal A},{\cal
A})} \times \int dU {\rm Tr}[U^n]\chi_{R}(U) \chi_{S}^{\dagger}(U),
\eeeq the notation being as in Eq.~\re{wilson}.  It can be shown that in the
decompactification limit $A\to\infty$, $\AA$ fixed, the following
expression is recovered 
\beeq \nome{wilmorenice}
&&\Wn(\AA;N)=\frac{1}{n N} \, \exp\lq-\frac{g^2 \AA}4 \, n(N+n-1)\rq
\sum_{k=0}^{+\infty}\frac{(-1)^k}{k!}\frac{\Gamma(N+n-k)}
{\Gamma(N-k)\Gamma(n-k)}\nonumber\\ &&\ph{\Wn(\AA;N)=\frac{1}{n N} \,
\exp\lq-\frac{g^2 \AA}4 \, n\rq} \times \exp\lq\frac{g^2
\AA}2\,n\,k\rq\,.  \eeeq 
The series is actually a finite sum, stopping at $k=n-1$ or $k=N-1$,
depending on the smaller one.  As happened for $n=1$,
Eq.~\re{wilmorenice} is expected to come out from the resummation of
't Hooft perturbative series, that corresponds to a light-front
quantization of the theory.  Some comments are now in order.  First of
all notice that when $n>1$ the simple abelian-like exponentiation is
lost. In other words the theory starts to feel its non-abelian nature
and the combinatorial coefficients in Eq.~\re{wilmorenice} signal that
the light-front vacuum, though simpler than its analog in the
equal-time quantization, cannot be considered trivial.  Actually, from
the sphere point of view, Eq.~\re{wilmorenice} can be understood as
coming from an instantons' resummation.  The procedure to see this is
exactly the one outlined above for the case $n=1$.  On the other hand
to neglect instantons, and then to send the area of the sphere to
infinity, is likely to reproduce the WML computation. In fact the
perturbative analysis have confirmed these claims.  Furthermore,
Eq.~\re{wilmorenice} exhibits an intriguing duality
\beq \nome{dual}
\Wn (\AA;N)=\WW_N \lp\frac nN \AA; n\rp\,, 
\eeq 
a relation that is far
from being trivial, involving an unexpected interplay between the
geometrical and the algebraic structure of the theory.  Looking at
Eq.~\re{dual}, the abelian-like exponentiation for $U(N)$ when $n=1$
appears to be connected to the $U(1)$ loop with $N$ windings, the
``genuine'' triviality of Maxwell theory providing the expected
behaviour for the string tension. Finally, one should observe that the
large-$N$ limit (with $n$ fixed) is equivalent to the limit in which
an infinite number of windings is considered with vanishing rescaled
loop area.  Alternatively, this rescaling could be thought to affect
the coupling constant $g^2 \to \frac{n}{N} g^2$. In detail, the former
limit ($N\to\infty$, ${\hat{g}}^2=g^2 N$ fixed) coincides with the
Kazakov-Kostov result~\cite{kazakov} 
\beq
\nome{old} 
\Wn(\AA;\infty) =
\frac1n L^{(1)}_{n-1}\lp \hat{g}^2 \AA\,n /2\rp \, \exp \lq
-\frac{\hat{g}^2 \AA\, n}4 \rq \,.  \eeq 
Next, using Eq.~\re{dual} one is able
to perform the latter limit, namely $n\to \infty$ with fixed
$n^2\,\AA$ 
\beq 
\nome{granden} 
\lim_{n\to\infty} \Wn(\AA;N) = \frac 1N
\, L^{(1)}_{N-1}\lp g^2 \AA\,n^2 /2 \rp \, \exp \lq -\frac{g^2 \AA\,
n^2}4\rq \,.  
\eeq 
Eq.~\re{granden} turns out to be  {\em exactly} the zero-instanton
contribution ${\cal W}^{(0)}_n$  in $\Wn (A-\AA,\AA)$ after
decompactification~\cite{bgv}. Such a coincidence can be explained by
observing  that, having the instantons a
finite size, small loops are essentially blind to them. 
Again, this is not the end of the story: as expected, Eq.~\re{granden} can be
derived through resummation of the perturbative series defined via WML
prescription (matrix model).

The conclusion to be drawn is that the interpretation of both the
equal-time and the light-front vacua in the case of a simple Wilson
loop can be extended to the more general case of a loop with winding
number $n$. In the light of the considerations above, WML, even
resummed, appears to be truly perturbative since it provides only the
expression of the zero-instanton contribution to Wilson loops
$\WW^{(0)}$. As opposed to this, 't Hooft is non-perturbative, in the
sense that a ``perturbative'' series in a ``static'' potential
reproduces a complex instanton expansion.
This seems to be the case also for loops in the adjoint representation
with non trivial $\th$-vacua (when the theory is based on the group
$SU(N)/\mathbb{Z}_N$). Again $\WW^{(0)}$ corresponds to a matrix model after
decompactification and is insensitive to the choice of the vacuum
sector. 
This will be the subject of a forthcoming publication.

\vfill\eject

\end{document}